\documentstyle[11pt,epsf]{article}
\newcommand{\unify}  {plot1}    
\newcommand{\mhvsmt} {plot7}    
\newcommand{\bq}{\begin{equation}}
\newcommand{\eq}{\end{equation}}
\newcommand{\rG}   {{\rm GUT}}
\newcommand{\MG}   {{\ifmmode M_\rG         \else $M_\rG$          \fi}}
\newcommand{\mb}   {{\ifmmode m_{b}         \else $m_{b}$          \fi}}
\newcommand{\mt}   {{\ifmmode m_{t}         \else $m_{t}$          \fi}}
\newcommand{\agut} {{\ifmmode \alpha_\rG    \else $\alpha_\rG$     \fi}}
\newcommand{\mgut} {{\ifmmode M_\rG         \else $M_\rG$          \fi}}
\newcommand{\mze}  {{\ifmmode m_0           \else $m_0$            \fi}}
\newcommand{\mha}  {{\ifmmode m_{1/2}       \else $m_{1/2}$        \fi}}
\newcommand{\tb}   {{\ifmmode \tan\beta     \else $\tan\beta$      \fi}}
\newcommand{\mz}   {{\ifmmode M_{Z}         \else $M_{Z}$          \fi}}
\newcommand{\ai}   {{\ifmmode \alpha_i      \else $\alpha_i$       \fi}}
\newcommand{\aii}  {{\ifmmode \alpha_i^{-1} \else $\alpha_i^{-1}$  \fi}}
\newcommand{\DRbar}{{\ifmmode \overline{DR} \else $ \overline{DR}$ \fi}}
\newcommand{\msusy}{{\ifmmode M_{SUSY}      \else $M_{SUSY}$       \fi}}
\newcommand{\as}   {{\ifmmode \alpha_s      \else $\alpha_s$       \fi}}
\newcommand{\asmz} {{\ifmmode \alpha_s(M_Z) \else $\alpha_s(M_Z)$  \fi}}
\newcommand{\tal}  {{\ifmmode \tilde{\alpha}\else $\tilde{\alpha}  \fi}}
\newcommand{\rb}[1]{\raisebox{1.5ex}[-1.5ex]{#1}}

\newcommand{\sws}  {{\ifmmode \;\sin^2\theta_W
                     \else    $\;\sin^{2}\theta_{W}$               \fi}}
\newcommand{\cws}  {{\ifmmode \;\cos^2\theta_W
                     \else    $\;\cos^{2}\theta_{W}$               \fi}}
\newcommand{\sw}   {{\ifmmode\;\sin\theta_W\else $\sin\theta_{W}$  \fi}}
\newcommand{\cw}   {{\ifmmode\;\cos\theta_W\else $\;\cos\theta_{W}$\fi}}
\newcommand{\tw}   {{\ifmmode\;\tan\theta_W\else $\;\tan\theta_{W}$\fi}}
\newcommand{\nn}   {\nonumber \\}

\begin{document}
%
%
\begin{titlepage}
\begin{flushright}
\vspace*{-2.7cm}
        IEKP-KA/94-07    \\
        May, 1994        \\
\end{flushright}
\vspace{1.5cm}
\begin{center} {\Large\bf Is the stop mass below the top mass?  \\}
\vspace{1.0cm}
{\bf W.  de Boer\footnote{Bitnet:
DEBOERW@CERNVM} and R. Ehret\footnote{Bitnet: BD21@DKAUNI2} \\} {\it
Inst.\ f\"ur Experimentelle Kernphysik, Univ.\ of Karlsruhe   \\} {\it
Postfach 6980, D-76128 Karlsruhe 1, FRG                         \\} and
\\ {\bf D.I. Kazakov\footnote{E-mail: KAZAKOVD@THSUN1.JINR.DUBNA.SU} \\}
{\it Lab. of Theoretical Physics, Joint Inst. for Nuclear Research, \\}
{\it 141 980 Dubna, Moscow Region, RUSSIA \\}
\end{center}

\vspace{2cm}

\begin{center}
{\bf Abstract}
\end{center}

\vspace{0.3cm}

\begin{center}\parbox{13cm}{\small

It is shown that a top mass  of $174\pm17$ GeV, as quoted  recently
by the CDF Collaboration, constrains the mixing angle between the Higgs
doublets in the Minimal Supersymmmetric extension of the Standard Model
(MSSM) to: $1.2<\tb<5.5$ at the 90\% C.L.. The most
probable value corresponds to $\tb = 1.56$; such a small value causes a
large mixing in the stop sector and the lightest stop is likely to be
below the top mass. In this case the stop production in $p\bar{p}$
collisions would contribute to the top signature, thus
providing a possible  explanation for the large effective
$t\bar{t}$ cross section observed by CDF.
             }
\end{center}
\vspace{0.5cm}



\end{titlepage}
%
%
\section{Introduction}

The Grand Unification idea has been  subjected  recently
to a new test using the new precise LEP data~\cite{ekn,abf,lalu}. The
result clearly indicates that the minimal Standard Model (SM) does not lead to
unification of the coupling constants, if they are extrapolated to
high energies~\cite{abf}.   On the contrary, within the  Minimal
Supersymmetric    extension of the
Standard Model (MSSM) unification is achieved.
Supersymmetry~\cite{rev} presupposes a symmetry between fermions
and bosons, thus doubling the particle spectrum of the
SM. The predicted particles are indicated by a tilde
above the usual SM symbol. Since these supersymmetric
particles (``sparticles'') have not yet been observed,
supersymmetry must be broken.
 From
the unification condition a first estimate of the SUSY breaking scale
could be made: it was found to be of the order of 1000 GeV, or more
precisely $10^{3\pm1}$  GeV~\cite{abf}.

Assuming soft symmetry breaking at the Grand Unification
 (GUT) scale, all    sparticle masses  can be expressed in terms of 5
parameters and the masses at low energies are
then determined by the well known Renormalization Group (RG) equations.
The parameters are: $m_0$, the common  mass   of the  spin 0 squarks and
sleptons;
$m_{1/2}$, the common mass of the spin 1/2 gauginos;
$\mu$, the mixing parameter between the Higgs doublets;
$\tb$, the ratio of the vacuum expectation values of the
two Higgs doublets; and $A$, the trilinear coupling in the Higgs sector.
So many parameters cannot be derived from the unification condition
alone. Further constraints can be considered:
\begin{itemize}
  \item $M_Z$ predicted from electroweak  symmetry
        breaking~\cite{Ino1} --
        \nocite{ewbr,ehnt83,grz90,Ibanez,rrb92,ir92,roskane}
        \cite{loopewbr}.
  \item Constraints from the unification of Yukawa
        couplings~\cite{rrb92,ir92},~\cite{Ross1} --
        \nocite{copw,bbo,bbog,lanpol,bmaskln,cpr,op,copw,cpw}
        \cite{ara91}.
  \item Constraints from the lower limit on the proton
        lifetime~\cite{langac,arn,protkln}.
  \item Experimental lower limits on SUSY masses~\cite{higgslim,pdb}.
  \item Constraints from the top mass suggested by CDF~\cite{cdf}.
\end{itemize}

We perform  a statistical
analysis, in which all constraints are
implemented in a $\chi^2$ definition and try to
find the most probable region of the parameter
space by minimizing the $\chi^2$ function.
The results   will be
presented after a short description
of the experimental input values and it is shown that   a likely  solution
has the lightest stop mass  below the top mass.

%

\section{Unification of the Couplings}
In the SM based on the group $\rm SU(3)\times
SU(2)\times U(1)$   the
 couplings are defined as:
\bq\label{SMcoup}{\matrix{
\alpha_1&=&(5/3)g^{\prime2}/(4\pi)&=&5\alpha/(3\cos^2\theta_W)\cr
\alpha_2&=&\hfill g^2/(4\pi)&=&\alpha/\sin^2\theta_W\hfill\cr
\alpha_3&=&\hfill g_s^2/(4\pi)\cr}}
\eq%
where $g'~,g$ and $g_s$ are the $U(1)$, $SU(2)$ and $SU(3)$ coupling
constants; the first two coupling constants are related to the fine
structure constant by:
$
  e = \sqrt{4\pi\alpha}=g\sin\theta_W=g'\cos \theta_W.
$

In the $\overline{MS}$ renormalization scheme the world averaged values
of the coupling constants  at the
Z$^0$ energy are
\begin{eqnarray}
  \label{worave}
  \alpha^{-1}(M_Z)             & = & 127.9\pm0.1\\
  \sin^2\theta_{\overline{MS}} & = & 0.2324\pm0.0005\\
  \alpha_3                     & = & 0.123\pm0.006.
\end{eqnarray}
The value of $\alpha^{-1}$ is given in ref.~\cite{dfs} and the value
of $\sin^2\theta_{\overline{MS}}$ has been  been taken from a detailed
analysis of all available data by
Langacker and Polonsky~\cite{sinms2}, which agrees
with the latest analysis of the LEP data~\cite{lep}.
The error includes the uncertainty
from the top quark. We have not used the smaller
error of 0.0003 for a given value of $\mt$, since
the fit was only done within the SM, not the MSSM,
so we prefer to use the more conservative error including
the uncertainty from $\mt$.

The $\alpha_3$ value
corresponds to the value at \mz\ as determined from quantities
calculated in the ``Next to Leading Log Approximation''~\cite{asresum}.
These quantities are less sensitive to the renormalization scale,
which is an indicator of the unknown higher order corrections;
they are the dominant uncertainties in quantities relying on second
order QCD calculations~\cite{asrev}. This $\alpha_s$ value is in
excellent agreement with a preliminary value of $0.120\pm 0.006$
from a fit to the $Z^0$ cross sections and
asymmetries measured at LEP~\cite{lep}, for which   the
third order QCD corrections have been calculated too;
the renormalization scale uncertainty is correspondingly small.

The top quark mass was simultaneously fitted to all electroweak data
and found to be~\cite{lep}:
\bq M_{top}=166^{+17~+19}_{-19~-22}~{\rm GeV},\label{mtop}\eq
where the first error is statistical and the
second error corresponds to a variation
of the Higgs mass between 60 and 1000 GeV.
The central value  corresponds to a Higgs mass of 300 GeV. Preliminary
analysis including the LEP 1993 data
find $M_{top}=165\pm 12\pm 18$ GeV and $M_{top}=174\pm 11^{+17}_{-19}$
GeV, if the new SLD data from SLAC is included\cite{alt}.
These values are  in good agreement with recent results  quoted by
the CDF Collaboration~\cite{cdf}:
\bq M_{top}=174^{+10~+13}_{-10~-12}~{\rm GeV},\label{mtop1}\eq
where the first error is statistical and the
second error systematic.

For SUSY models, the dimensional reduction $\overline{DR}$
scheme is a more appropriate renormalization scheme~\cite{akt}.
This scheme also has the advantage
that all thresholds can be treated by simple step
approximations.  Thus
unification occurs in the $\overline{DR}$ scheme if all
three $\aii(\mu)$ meet
exactly at one point.
                     This crossing point then gives
the mass of the heavy
gauge bosons.  The $\overline{MS}$ and $\overline{DR}$ couplings
differ by a small offset
\bq\label{MSDR}{{1\over\alpha_i^{\overline{DR}}}=
{1\over\alpha_i^{\overline{MS}}}-{C_i\over\strut12\pi}
}\eq
where the $C_i$ are the quadratic Casimir coefficients
of the group ($C_i=N$
for SU($N$) and 0 for U(1) so $\alpha_1$ stays the same).
 Throughout the
following, we use the $\overline{DR}$ scheme for the MSSM.

\section{$M_Z$    from Electroweak Symmetry Breaking}

In the MSSM at least two Higgs doublets have   to be introduced.
Radiative corrections from the heavy top and stop quarks can drive
one of the Higgs masses negative, thus causing spontaneous symmetry
breaking in the electroweak sector. In this case the Higgs potential
does not have its minimum for all fields equal zero, but the
minimum is obtained for non-zero vacuum expectation  values of the
fields.  The scale, where symmetry breaking occurs depends on the
 starting values of the mass parameters at the GUT scale,
the top mass and the evolution of the couplings and masses.
This gives strong constraints between the known $Z^0$ mass and the
SUSY mass parameters, as demonstrated e.g. in ref.~\cite{rrb92}.

After including  the one-loop corrections to the
potential~\cite{erz,berz,drno,kz92,eqz,cpr},
the $M_Z$ mass becomes dependent on the top- and stop quark masses too.
The corrections   are zero if the top- and stop
quark masses are identical, i.e. if supersymmetry would be exact.
They grow with the difference $\tilde{m}^2_t-m_t^2$, so these
corrections become unnaturally large for large values of the
stop masses \cite{bek1,wdb}.

\section{$m_b$ from the $m_b/m_\tau$ Mass Ratio  }

Unification of the Yukawa couplings for a given generation at the
GUT scale predicts relations for quark and lepton masses within a given
family. Unfortunately, for the light quarks  the masses  are uncertain,
but the ratio of b-quark and $\tau$-lepton masses can be correctly
predicted by the radiative  mass
corrections~\cite{bmas,Ibanez,lanpol,bmaskln,bbo,bbog}.

Assuming the simplest possible GUT model based on SU(5) gauge group, one
has at the GUT scale: $m_b = m_\tau $. To calculate the
experimentally observed mass ratio the RG equations for the
running masses have to be used. By a physical
mass we understand the value of the running mass at the energy scale
equal to the mass itself. This definition of the mass is used
throughout this paper.

 From the RG equations for the Yukawa couplings one can easily
obtain the RGE  for the ratio \cite{bek1,wdb}
$$R_{b\tau } \equiv \frac{m_b}{m_\tau } = \sqrt{\frac{Y_b}{Y_\tau }}.$$

For the running mass of the b-quark we used~\cite{runmas}:
\bq m_b=4.25\pm0.3~ {\rm GeV}.\label{bmas}\eq
This mass depends on the choice of scale
  and the value of $\as(m_b)$.
Consequently, we have assigned a rather
conservative error of 0.3 GeV instead of
the proposed value of 0.1 GeV~\cite{runmas}.
Note that the running mass (in the
$\overline{MS}$ scheme) is related to the
physical (pole) mass $M_b^{pole}$ by~\cite{runmas}:
\bq m_b=M_b^{pole}\left(1-\frac{4}{3}\frac{\as}
{\pi}-12.4(\frac{\as}{\pi})^2\right)\approx 0.825 \;M_b^{pole}, \eq
so $m_b=4.25$ corresponds to $M_b^{pole}\approx 5$ GeV.
We ignore the running of $m_\tau $ below
$m_b$ and use for the $\tau$ mass:
$M_\tau=1.7771\pm 0.0005$ GeV~\cite{taumas}.

\section{Top Mass Constraints}
\label{s65}
The top mass can be expressed as:
\bq
  \mt^2=(4\pi)^2\ Y_t(t)\ v^2\ \sin^2(\beta), \label{mt}
\eq
where the running of the Yukawa coupling $Y_t$ as function of
$t=\log(\frac{M_\rG^2}{Q^2})$ in first order\footnote{Throughout the
analysis we have used the second order RG equations, for which
no analytical solution exists, but this will not
change the following arguments dramatically.}
is given by~\cite{Ibanez}:
\bq
  Y_t(t)=\frac{\displaystyle Y_t(0)E(t)}{\displaystyle 1+6Y_t(0)F(t)},
  \label{ytt}
\eq
where $E$ and $F$ are functions of the couplings only~(see appendix).
One observes that $Y_t(t)$ becomes independent
of $Y_t(0)$ for large values of $Y_t(0)$, implying
an upper limit on the top mass.
 Requiring electroweak symmetry breaking
implies a minimal value of the top Yukawa
coupling, typically $Y_t(0)\ge {\cal O}(10^{-2})$.
In this case the term
   $6Y_t(0)F(t)$ in the denominator of eq.~(\ref{ytt})
is much larger than one, since $F(t)\approx 290$ at
the weak scale, where  $t\approx 66$.
 In this case $Y_t(t)=E(t)/6F(t)$, so from eq.~(\ref{mt}) it follows:
\bq m_t^{2}=\frac{(4\pi)^2\ E(t)}{6F(t)}\ v^2\ \sin^2(\beta)\approx
(190~{\rm GeV})^2\sin^2(\beta),\eq
The physical (pole) mass is about 6\% larger
than the running mass~\cite{runmas}:
\bq M_{t}^{pole}=m_t\left(1+\frac{4}{3}\frac{\as}
{\pi}\right)\approx (200~{\rm GeV})\sin\beta,\label{topm}.\eq

The electroweak breaking conditions
require $\pi/4<\beta<\pi/2$ ; hence
the equation above implies for the MSSM approximately:
\bq 145 < M_{t}^{pole} < 200 ~{\rm GeV},\label{mtlim} \eq
which is consistent with
the experimental values given in eqns.~(\ref{mtop}) and (\ref{mtop1}).

For   large top masses, the b-quark
mass becomes a sensitive function of $\mt$
and of the starting values of the gauge
couplings at $M_\rG$ \cite{ir92}.
\section{Experimental Lower Limits on SUSY Masses}
SUSY particles have not been found so far
and from the searches
at LEP one knows that the lower limit on the
charged leptons and charginos is
about half the $Z^0$ ~ mass (45 GeV)~\cite{pdb}
and the Higgs mass has to be above
62 GeV~\cite{higgslim}. The lower limit on
the lightest neutralino is 18.4 GeV~\cite{pdb},
while the sneutrinos have to
be above 41 GeV~\cite{pdb}.
  These limits require  minimal values for the
SUSY mass parameters.

There exist also limits on squark and gluino
masses from the hadron colliders~\cite{pdb}, but these
limits depend on the assumed decay modes.
Furthermore, if one takes the limits given above
into account, the  constraints from the limits of all other
particles are usually fulfilled, so they
do not provide additional reductions of  the
parameter space in case of the {\it minimal} SUSY model.

\section{Proton Lifetime Limits}

GUT's predict proton decay and the present lower limits
on the proton lifetime yield quite strong  constraints
on the GUT scale and the SUSY parameters.
The direct decay $p\rightarrow e^+\pi^0$
via s-channel exchange requires
the GUT scale to be above $10^{15}$ GeV. This is not fulfilled
in the SM, but always fulfilled in the MSSM. Therefore we do not
consider this constraint.
However, the decay via box diagrams with winos and Higgsinos
predict much shorter lifetimes, especially in the preferred mode
 $p\rightarrow \overline{\nu} K^+$.
 From the present experimental lower limit of  $10^{32}$ yr for
this decay mode Arnowitt and Nath~\cite{arn}  deduce an upper limit
on the parameter B:
\bq
   B <\; (293\pm 42)\;M_{H_3}/3M_\rG\ GeV^{-1}
\eq
Here $M_{H_3}$ is the Higgs triplet mass, which is expected to be
of the order of $M_\rG$. To obtain a conservative upper limit
on $B$, we allow  $M_{H_3}$ to become an order of magnitude heavier
than $M_\rG$, so we require
\bq
  B< 977\pm 140\ GeV^{-1}.
\eq
The uncertainties from the unknown heavy Higgs mass  are large
compared with the contributions from the first and third generation,
which contribute through the mixing in the CKM matrix.
Therefore we only consider the second order generation
contribution, which can be written as~\cite{arn}:
\bq
  B=\frac{-2 \alpha_2}{ \alpha_3  \sin(2\beta)}
  \frac{ m_{\tilde{g}}}{ m^2_{\tilde{q}}} ~10^6
  \label{prot}
\eq
where $\alpha_2$ and $\alpha_3$ are the coupling
constant of the $SU(2)$ and $SU(3)$ groups at the SUSY scale,
respectively.
One observes that the upper limit on $B$ favours small gluino masses
$m_{\tilde{g}}$, large squark masses $ m_{\tilde{q}} $, and small
values of $\tb $.
 To fulfill this constraint requires
\bq
   \tan \beta < 10
\label{tanb}
\eq
for practically the whole parameter space                    \cite{arn,bek1}.

\section{Fit Strategy}\label{s61}

 From the five parameters in the MSSM plus the common coupling $\agut$
at the unification scale
  $\mgut$ one can determine  all other SUSY masses, the b-quark mass,
and $\mz $   by performing the complete evolution of the couplings  and masses
including all thresholds. Details can be found in \cite{bek1,wdb}.

The most probable parameter values were obtained  by minimizing
the following $\chi^2$ function\footnote{We use the MINUIT
program from F. James and M. Roos, {\em MINUIT Function Minimization
and Error Analysis\/}, CERN Program Library Long Writeup D506;
Release 92.1, from March 1992. Our $\chi^2$ has discontinuities
due to the  experimental bounds on various quanitities, which
become ``active''only for specific regions of the
parameter space. Consequently the derivatives are
not everywhere defined. The option SIMPLEX, which does
not rely on derivatives,  can be used to find
the monotonous region and the option MIGRAD to
optimize inside this region.}:
\begin{eqnarray} \chi^2 & = &
    {\sum_{i=1}^3\frac{(\aii(\mz)-\alpha^{-1}_{MSSM_i}(\mz))^2}
                      {\sigma_i^2}}                             \nn
 & &+\frac{(\mz-91.18)^2}{\sigma_Z^2}                           \nn
 & &+\frac{(\mb-4.25)^2}{\sigma_b^2}                            \nn
 & &+{\frac{(B  - 997)^2}{\sigma_B^2}} {(for ~B > 997)}         \nn
 & &+{\frac{(D(m1m2m3))^2}{\sigma_D^2}} {(for~ D > 0)}          \nn
 & &+{\frac{(\tilde{M}-\tilde{M}_{exp})^2}{\sigma_{\tilde{M}}^2}}
 {(for~\tilde{M} > \tilde{M}_{exp})}.                  \label{chi2}
\end{eqnarray}
The first term is the contribution of the
difference between the three
calculated and measured gauge coupling
constants at \mz~and  the following
two terms are the contributions from the
\mz-mass ~and \mb-mass  constraints.
The last three terms impose constraints from the proton
lifetime limits, from
electroweak symmetry
breaking, i.e. $D=V_H(v_1,v_2)-V_H(0,0) < 0$,
and from experimental lower limits on the SUSY masses.
The top mass, or equivalently, the top
Yukawa coupling enters sensitively into the
calculation of $\mb$ and $\mz$.
Instead of the top Yukawa coupling
one could have taken the top mass as a parameter.
However, if the couplings are evolved from $\mgut$
downwards, it is more convenient to run also the
Yukawa coupling downward, since the RG equations of the
gauge and Yukawa couplings form a set of coupled
differential equations in second order.
Once the Yukawa coupling is known at $\mgut$,
the top mass can be calculated at any scale.

The following errors were attributed:
$\sigma_i$ are the experimental errors in the
coupling constants,
as given above, $\sigma_b$=0.3 GeV,
$\sigma_B=140~ \mbox{GeV}^{-1}$, while  $\sigma_D$ and
$\sigma_{\tilde{M}}$ were set to 10 GeV.
 The values of the  latter errors are not   critical,
 since the corresponding terms in the numerator
 are  zero
 in case of a good fit and even for the 90\% C.L.  limits
 these constraints could be fulfilled and the $\chi^2$ was determined
 by  the other terms, for which one knows the errors.

The light thresholds are taken into account in the evolution of the
coupling constants by
changing the coefficients of the RGE at the
value $Q=m_i$, where
the threshold masses $m_i$ are obtained from
the analytical solutions of the corresponding RGE.
These solutions depend on the integration range,
which was chosen between $m_i$ and $\mgut$.
However, since one does not know $m_i$ at the
beginning, an iterative procedure has to be used:
one  first uses $\mz$ as a lower integration limit,
calculates $m_i$,
and uses this as lower limit in the next iteration.
Of course, since the  coupling constants are running,
the latter have to be iterated too, so the values
of $\alpha_i(m_i)$ have to be used for calculating
the mass at the scale $m_i$~\cite{rrb92,acpz}.
Usually three to five iterations are enough to find a stable solution.

Following Ellis, Kelley and Nanopoulos~\cite{ekn2} the
possible effects from heavy thresholds are set to zero, since
the proton lifetime limits  forbid the Higgs triplet masses to be below \MG.
These heavy thresholds have been considered
by other authors   for different assumptions~\cite{baha,sinms2,acz}.

\section{Results}\label{s62}

\begin{figure}
 \begin{center}
  \leavevmode
  \epsfxsize=15cm
  \epsfysize=18cm
  \epsffile{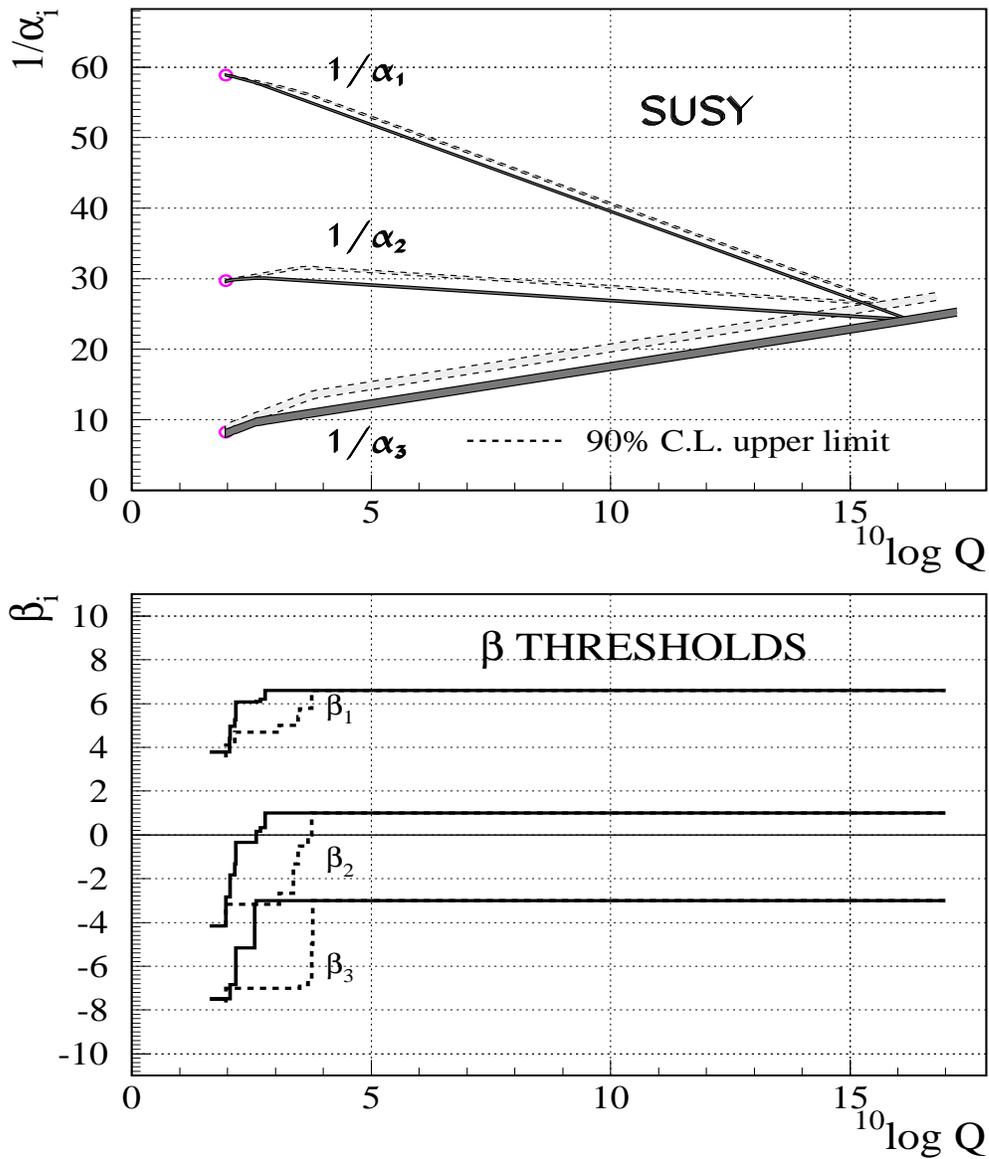}
 \end{center}
 \caption{\label{\unify}
 Evolution of the inverse of the three couplings in the
 MSSM.                   The line  above $\MG $ follows the prediction
 from the supersymmetric SU(5) model.
The SUSY thresholds have been indicated in the lower part  of the curve:
they are treated as step functions in the
     first order $\beta$ coefficients in the  renormalization group
equations, which correspond to a change in slope in the evolution
of the couplings in the top figure.
The dashed lines correspond to the 90\% C.L. upper limit for the
SUSY thresholds.
}
\end{figure}

 We first consider fits without proton lifetime  constraints, since
  they are only important in the determination of lower limits, as will
be discussed below.
The upper part of fig.~\ref{\unify}  shows the evolution of the
coupling constants in the MSSM for two cases: one for the minimum
value of the $\chi^2$ function given in eq.~\ref{chi2} (solid lines)
and one corresponding to the 90\% C.L. upper limit of the thresholds
of the light SUSY particles (dashed lines). The  position of the light
thresholds is shown in the bottom part as jumps in the first order
$\beta$ coefficients, which are increased   as soon as a new threshold is
passed.
Also the second order coefficients
are changed correspondingly,
but their effect on the evolution
is not visible in the top figure in contrast
to the first order effects, which change the
slope of the lines considerably in the top figure.
One observes that the changes  in the coupling constants
occur  in a rather
narrow energy regime, so qualitatively
this picture is very similar to the case,
in which all sparticles were assumed to be
degenerate at an  effective SUSY mass scale $M_{\rm SUSY}$~\cite{abf}.
Since the running of the couplings depends only
logarithmically on the sparticle masses, the
90\% C.L. upper limits are as large as several TeV, as
shown by the dashed lines in fig.~\ref{\unify} and more
quantitatively in table \ref{t71}.
With the fitted SUSY parameters given at the top of the table,
the corresponding masses of the SUSY particles
can be calculated. Their values are given in the
lower part of the table.

The 90\% C.L. upper and lower limits on the masses are obtained by
scanning $\mze$ and $\mha$ till
the $\chi^2$ value increases by 1.64, while optimizing the
values of $\tb,~ \mu,~,\agut,~Y_t(0)$ and $\mgut$. The lower
limits on the SUSY parameters are shown in the left column of
table \ref{t71}. The lowest values of $\mze$
and $\mha$ are required to have simultaneously a
sneutrino mass above 42 GeV and a wino  mass above 45 GeV.
If the proton lifetime limit is included, either $\mze$ or $\mha$ have
to be above a certain limit (see eq.  \ref{prot}).
Since the squarks and gauginos are much more sensitive to
$\mha$ than $\mze$, one obtains the lower limits by increasing $\mze$.
The minimum value for $\mze$ is about 400 GeV in this case.
But in both cases the $\chi^2$ increase for the
lower limits is  due to the b-mass,
which is predicted to be 4.6 GeV from the
parameters determining the lower limits, so it gives a contributions to
the $\chi^2$ function,
which requires $m_b$=4.25 GeV (see eq. \ref{bmas}).
The 90\% C.L. upper limits can be several TeV.
 If one requires that only
solutions are allowed for which the corrections to $\mz$
are not large compared to $\mz$ itself, one has to limit
the mass of the heaviest stop quark to about one TeV. The
corresponding 90\% C.L. upper limits of the individual
sparticles masses are given in the right hand column of table
\ref{t71}. The correction to $\mz$ is 6 times \mz~ in this case.

\begin{figure}
 \vspace{-2cm}
 \begin{center}
  \leavevmode
  \epsfxsize=13cm
  \epsfysize=16cm
  \epsffile{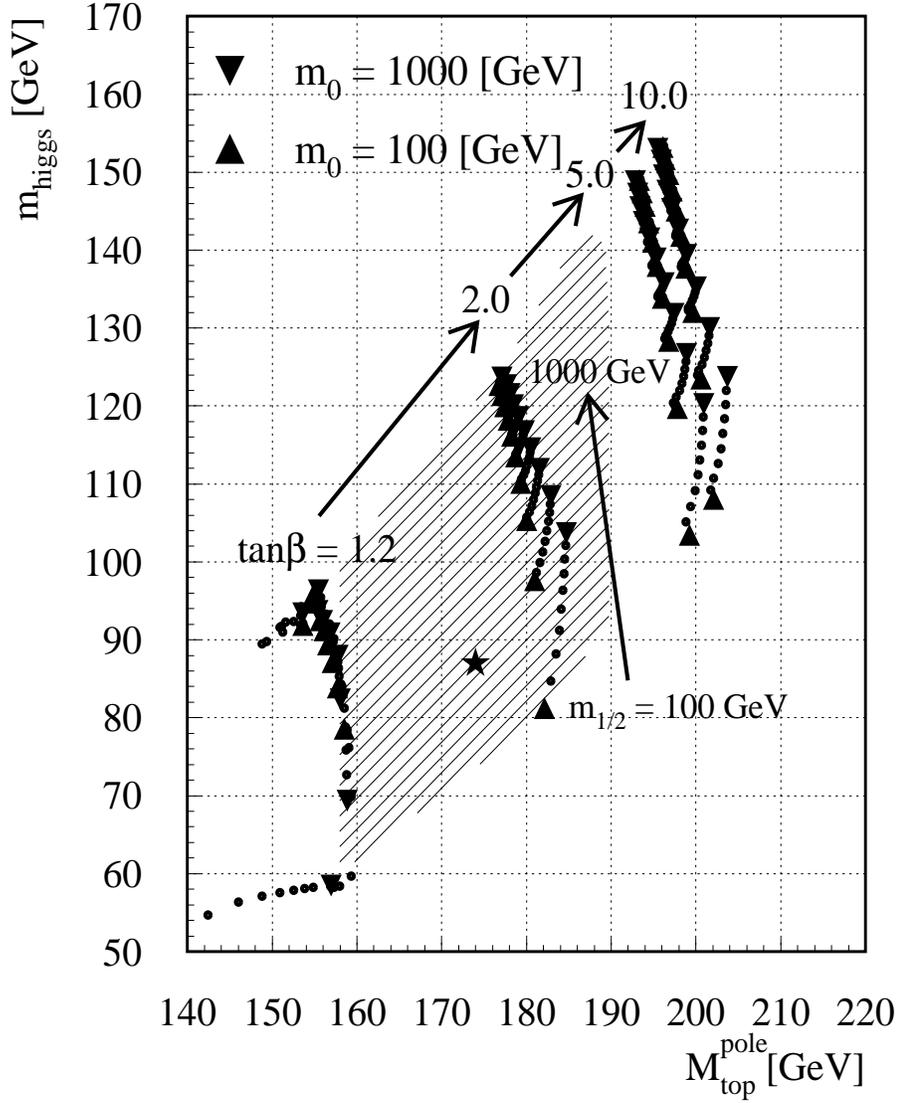}
 \end{center}
 \caption[ ]
  {The mass of the lightest Higgs particle as function of the
   top quark mass for values of $\tb$ between 1.2 and 10 and values
   of $\mze$ and $\mha$ between 100 and 1000 GeV.
   The parameters  of $\mu, ~\mgut,~ \agut$ and $Y_t(0)$~are optimized
   for each choice of these parameters; the corresponding values of
   the top and lightest Higgs mass  are shown as symbols.
   For small values of $\mha$ the Higgs mass increases with $\mze$,
   as shown for a ``string'' of points, each representing a step of
   100 GeV in $\mze$ for a given value of $\mha$, which is increasing
   in steps of 100 GeV, starting with the low values for the lowest
   strings. At high values of $\mha$ the value of $\mze$ becomes
   irrelevant and the ``string'' shrinks to a point. Note the strong
   positive correlation between $m_{higgs}$ and all other parameters:
   the highest value of the Higgs mass corresponds to the maximum
   values of the input parameters, i.e. $\tb=10$, $\mze=\mha=1000$ GeV;
   this value does not correspond to the minimum $\chi^2$. More likely
   values are: $m_{higgs}\approx 87$ GeV for $\mha=100$ GeV,
   $\mze=400$ GeV, $\mu=822$ GeV and $\tb=1.6$, as indicated by the
   star. The hatched area corresponds to the top mass range measured
   by~\cite{cdf}.
  }
\label{\mhvsmt}

\end{figure}

The mass  of the lightest Higgs particle, called $h$ in table \ref{t71},
is a rather strong function of \mt, as shown in fig.~\ref{\mhvsmt}
for various choices of $\tb$, $\mze$ and $\mha$. All other parameters
were optimized for these inputs and after the fit the values of the
Higgs and top mass were calculated and plotted. One observes that the
mass of the lightest Higgs particle varies between 60 and 150 GeV and
the top mass between 134 and 190  GeV. Furthermore,
it is evident
that $\tb$ almost uniquely determines
the value of $\mt$ (through eq. \ref{mt}), since even
if $\mha$ and $\mze$ are varied between 100
and 1000 GeV, one finds
practically the same $\mt$ for a given
$\tb$.  The value of $\mt$
varies between 134 and 190 GeV, if $\tb$
is varied between 1.2 and 10.
This range is in excellent agreement with
the estimates given in
eq.~\ref{mtlim}, if one takes into account that
$M_t^{pole}\approx 1.06 \mt$ (see eq.~\ref{topm}).
The shaded area in   fig.~\ref{\mhvsmt}  ~indicates
the results on the top mass quoted by the CDF Collaboration\cite{cdf}.
It clearly favours low values
of $\tb$.
Adding to the $\chi^2$ a term $(M_t-174)^2/17^2$ yields
after minimization:
\bq 1.2<\tb<5.5~{\rm at ~the~ 90\% ~C.L.} \eq
The most probable value corresponds to $\tb=1.56$ as indicated by the
star in the figure.

Such a low value leads to a large mixing in the stop sector, in which
case a likely  value of the lightest stop
is below the top mass (see the typical fit in table
\ref{t71}), although stop masses above the top mass
are not excluded, as shown by the upper limits in table
 \ref{t71}. Also a change in sign of  the Higgs mixing
parameter $\mu$ leads to stop masses above the top mass,
but   the $\chi^2$ value is hardly worse in that case,
so this cannot be excluded either.
Varying $A_t(0)$ between
$+3\mze$ and $-3\mze$
does not influence the results very much, since
it is usually compensated by a change in  $\mu$, so
 $A_t(0)$  was kept  zero,
but its non-zero value at lower energies      was taken
into account.

If the  stop mass is below the top mass, it
cannot decay into the top, but can decay as follows:
$$
  \tilde{t}_1\rightarrow \tilde{\chi}^\pm_1+b
  \rightarrow\tilde{\chi}^0_1+W+b \rightarrow
  \tilde{\chi}^0_1+lepton+\nu+b,
$$
which is experimentally very similar to the normal top decay
signature \cite{baer}.
Additional stop production could be an explanation for the excess of
events seen by the CDF Collaboration: they observe an effective cross
section for top pair production  of $13.9^{+6.1}_{-4.8}$ pb, while the
calculated $t\bar{t}$ cross section is only $5.8^{+0.8}_{-0.4}$
pb~\cite{cdf}.
\medskip\\
{\bf Acknowledgement.}
\medskip\\
We thank G.~Altarelli, M.~Drees and T.~Sj\"ostrand for discussions.

\begin{table}[h]
\vspace*{-4cm}
\renewcommand{\arraystretch}{1.30}
\renewcommand{\rb}[1]{\raisebox{1.75ex}[-1.75ex]{#1}}
\begin{center}
\begin{tabular}{|c|r|r||r||r|r|}
\hline
Symbol& \multicolumn{2}{|c||}{ \makebox[4.6cm]{Lower limits }}&
      \makebox[2.3cm]{{\bf Typical fit}} &
      \multicolumn{2}{|c|}{ \makebox[4.6cm]{90\% C.L.~Upper limits }} \\
\hline
\hline
Constraints & \makebox[2.3cm]{ GEY } &\makebox[2.3cm]{ GEY+P } &
 \makebox[2.3cm]{ {\bf  GEY+(PF) }}&
 \makebox[2.3cm]{ GEY+ (P)  }      &
 \makebox[2.3cm]{ GEY+(P)+F  }                                        \\
\hline
\hline
 \multicolumn{6}{|c|}{ Fitted SUSY parameters }                       \\
\hline
 $m_0$   &  65 &400&{\bf 400} &  400 & 400                            \\
\hline
 $m_{1/2}$  &37 &80 &{\bf 111} & 1600 & 475                           \\
\hline
 $\mu$  &-117&330&{\bf 870} & 1842 & 1101                             \\
\hline
 $\tan\beta$  &3.0 &3.0&{\bf 1.56} &  8.5& 2.9                        \\
\hline
 $Y_t(0)$   &0.0158 &0.0035&{\bf 0.0150} & 0.0023 & 0.0084            \\
\hline
$M_t^{pole}$   & -- & -- &{\bf 175} &  178 & 189                      \\
\hline
$m_t$   & -- & -- &{\bf 165} &  168 & 178                             \\
\hline
 $1/\alpha_\rG$  &23.8&24.3&{\bf 24.5} & 25.9 & 25.2                  \\
\hline
 $\MG$  &$2.3\;10^{16}$&$2.0\;10^{16}$& $
 {\bf 2.0\;10^{16}}$ & $0.8\;10^{16}$ & $1.3\;10^{16}$                \\
\hline
\hline
 \multicolumn{6}{|c|}{SUSY masses in [GeV]}                           \\
\hline
\hline
  $\chi^0_1(\tilde{\gamma})$   & 18 &25 &{\bf  41}& 720  &  202       \\
\hline
  $\chi^0_2(\tilde{Z})$  &39&52&{\bf  80}  & 1346  &  386             \\
\hline
  $\chi^{\pm}_1(\tilde{W})$   &  46&48&{\bf  79}  & 1347 &  386       \\
\hline
  $\tilde{g}$    &  109 &217&{\bf 293}& 3377 & 1105                   \\
\hline  \hline
  $\tilde{e}_L$      &  82 &406&{\bf 409}& 1160  &  521               \\
\hline
  $\tilde{e}_R$    &  67&401&{\bf 402} & 729  &  440                  \\
\hline
  $\tilde{\nu}_L$     &  41&400&{\bf406}  & 1157  &  516              \\
\hline  \hline
  $\tilde{q}_L$   &  120&443&{\bf 477} & 3030  & 1071                 \\
\hline
  $\tilde{q}_R$   &  115&440&{\bf 471} & 2872  & 1030                 \\
\hline
 $\tilde{b}_L$   &  112&352&{\bf 369} & 2610  & 903                   \\
\hline
  $\tilde{b}_R$   &  119&440&{\bf 471} & 2862  & 1027                 \\
\hline
  $\tilde{t}_1$    & --& --&{\bf 144} & 2333  &  725                  \\
\hline
  $\tilde{t}_2$  &  -- &  --&{\bf 467} & 2817  & 1008                 \\
\hline        \hline
  $ \chi^0_3(\tilde{H}_1)$  &  109  & 292&{\bf  540}& 1771  &  799    \\
\hline
  $\chi^0_4(\tilde{H}_2)$   &  120&313&{\bf 556}& 1780  &  812        \\
\hline
  $\chi^{\pm}_2(\tilde{H}^{\pm})$& 129&315&{\bf 566}& 1816  &  831    \\
\hline   \hline
  $       h $    &  --  &--&{\bf  87}&  146 &  127                    \\
\hline
  $       H $   & 118&523&{\bf 812}& 2218  &  1033                    \\
\hline
  $       A $   & 92 &521&{\bf 810}& 2217  &  1031                    \\
\hline
  $       H ^{\pm}$    &121&527&{\bf 813 }& 2219  &  1034             \\
\hline
\end{tabular} \end{center}
\caption[]{\label{t71} Values of SUSY masses and parameters for
  various constraints: G=gauge coupling unification;
  E=electroweak symmetry breaking;
  Y=Yukawa coupling unification;
  P=Proton lifetime constraint;
  F=finetuning constraint. Constraints in brackets indicate that they
  are fulfilled but not required. The minimum values of the lightest
  Higgs mass, the stop mass and the top mass can't be reached
  for the parameters minimizing the squarks and slepton masses.
  One needs smaller values of $\tb$ in that case.}
\end{table}

%

\clearpage
%
%
%
%


\end{document}